\documentclass[sigplan,screen]{acmart}
\DeclareUnicodeCharacter{2032}{\ensuremath{\prime}}

\AtBeginDocument{%
  }

\setcopyright{none}
\acmYear{2025}
\acmConference[EGRAPHS '25]{EGRAPHS Workshop}{June 17,
  2025}{Seoul, South Korea}

\usepackage[utf8]{inputenc}
\usepackage[T1]{fontenc}
\usepackage{microtype}

\usepackage{xargs}
\usepackage{xparse}
\usepackage{xifthen, xstring}
\usepackage{xspace}
\usepackage{marginnote}
\usepackage{etoolbox}
\usepackage{amsmath}
\usepackage{thmtools} 
\usepackage{algorithm}
\usepackage[noend]{algpseudocode}
\usepackage{listings}
\usepackage{fancyvrb}
\usepackage{acro}

\usepackage{tikz}
\usetikzlibrary{fadings}
\definecolor{pagebg}{rgb}{0.95,0.95,0.95} 

\colorlet{fadecolor}{pagebg} 

\tikzfading[name=fade out bottom,
            top color=transparent!0,
            bottom color=transparent!100]

\usepackage[textsize=tiny]{todonotes}
\usepackage[normalem]{ulem}

\makeatletter
\font\uwavefont=lasyb10 scaled 652
\newcommand\colorwave[1][blue]{\bgroup\markoverwith{\lower3\p@\hbox{\uwavefont\textcolor{#1}{\char58}}}\ULon}

\newcommand\highlight[2]{{\color{#1}{\colorwave[#1]{#2}}}}
\makeatother

\makeatletter
\newcommand\InFloat[2]{\ifnum\@floatpenalty<0\relax#1\else#2\fi}
\makeatother

\newboolean{inComment}
\setboolean{inComment}{false}

\ifx\paperversion\paperversiondraft
\newcommand\createtodoauthor[2]{
  \def\tmpdefault{emptystring}
  \expandafter\newcommand\csname #1\endcsname[2][\tmpdefault]{
    \ifthenelse{\boolean{inComment}}{
      \PackageError{paper-template}{Comments in comments not supported}{}
    }{}\setboolean{inComment}{true}
    \def\tmp{##1}
    \InFloat{
        \smash{
	  \marginnote{
	    \todo[inline,linecolor=#2,backgroundcolor=#2,bordercolor=#2]
	      {\textbf{#1 (Figure):} ##2}
          }
        }
    }{\ifthenelse{\equal{\tmp}{\tmpdefault}} 
      {\todo[linecolor=#2,backgroundcolor=#2,bordercolor=#2]{\textbf{#1:} ##2}\ignorespaces}
      {\ifthenelse{\equal{##2}{}} 
        {\highlight{#2}{##1}}
        {\highlight{#2}{##1}\todo[linecolor=#2,backgroundcolor=#2,bordercolor=#2]
	  {\textbf{#1:} ##2}
	}
      }
    }
    \setboolean{inComment}{false}
  }
}
\else
\newcommand\createtodoauthor[2]{%
\expandafter\newcommand\csname #1\endcsname[2][]{##1}%
}%
\fi

\usepackage[frozencache,cachedir=.]{minted}
\usepackage{etoolbox}

\makeatletter
\ifcsdef{minted@optlistcl@quote}
{
\ifwindows
  \renewcommand{\minted@optlistcl@quote}[2]{%
    \ifstrempty{#2}{\detokenize{#1}}{\detokenize{#1="#2"}}}
\else
  \renewcommand{\minted@optlistcl@quote}[2]{%
    \ifstrempty{#2}{\detokenize{#1}}{\detokenize{#1='#2'}}}
\fi

\newcommand{\minted@def@optcl@novalue}[2]{%
  \define@booleankey{minted@opt@g}{#1}%
    {\minted@addto@optlistcl{\minted@optlistcl@g}{#2}{}%
     \@namedef{minted@opt@g:#1}{true}}
    {\@namedef{minted@opt@g:#1}{false}}
  \define@booleankey{minted@opt@g@i}{#1}%
    {\minted@addto@optlistcl{\minted@optlistcl@g@i}{#2}{}%
     \@namedef{minted@opt@g@i:#1}{true}}
    {\@namedef{minted@opt@g@i:#1}{false}}
  \define@booleankey{minted@opt@lang}{#1}%
    {\minted@addto@optlistcl@lang{minted@optlistcl@lang\minted@lang}{#2}{}%
     \@namedef{minted@opt@lang\minted@lang:#1}{true}}
    {\@namedef{minted@opt@lang\minted@lang:#1}{false}}
  \define@booleankey{minted@opt@lang@i}{#1}%
    {\minted@addto@optlistcl@lang{minted@optlistcl@lang\minted@lang @i}{#2}{}%
     \@namedef{minted@opt@lang\minted@lang @i:#1}{true}}
    {\@namedef{minted@opt@lang\minted@lang @i:#1}{false}}
  \define@booleankey{minted@opt@cmd}{#1}%
      {\minted@addto@optlistcl{\minted@optlistcl@cmd}{#2}{}%
        \@namedef{minted@opt@cmd:#1}{true}}
      {\@namedef{minted@opt@cmd:#1}{false}}
}
\minted@def@optcl@novalue{customlexer}{-x}
}
{
}
\makeatother

\usepackage{tikz}
\usetikzlibrary{arrows}
\usetikzlibrary{shapes}

\tikzset{
  circledstyle/.style={
    shape=circle,
    #1,
    font=\tt\small,
    inner sep=0pt,
    outer sep=0pt,
    minimum size=1.2em,
    text=black
  }
}

\usepackage[norefs,nocites]{refcheck}

\makeatletter
\newcommand{\refcheckize}[1]{%
  \expandafter\let\csname @@\string#1\endcsname#1%
  \expandafter\DeclareRobustCommand\csname relax\string#1\endcsname[1]{%
    \csname @@\string#1\endcsname{##1}\wrtusdrf{##1}}%
  \expandafter\let\expandafter#1\csname relax\string#1\endcsname
}
\makeatother

\refcheckize{\autoref}

\acmConference[EGRAPHS '25]{EGRAPHS Workshop}{June 17,
  2025}{Seoul, South Korea}
\acmBooktitle{EGRAPHS '25,
 June 17, 2025, Seoul, South Korea}
  \acmYear{2025}
  \acmISBN{}
  \acmDOI{}
  \startPage{1}

\setcopyright{none}             

\usepackage{minted}

\makeatletter
\ifcsdef{MintedExecutable}
{
  \newminted[mlir]{tools/lexers/MLIRLexer.py:MLIRLexerOnlyOps}{mathescape, breaklines,escapeinside=\#\%}
  \newminted[xdsl]{tools/lexers/MLIRLexer.py:MLIRLexer}{mathescape, style=murphy}
  \newminted[lean4]{tools/lexers/Lean4Lexer.py:Lean4Lexer}{mathescape}
}
{
  \ifcsdef{minted@optlistcl@quote}
  {
    \newminted[mlir]{tools/lexers/MLIRLexer.py:MLIRLexerOnlyOps}{customlexer, mathescape, breaklines, escapeinside=||}
    \newminted[xdsl]{tools/lexers/MLIRLexer.py:MLIRLexer}{customlexer, mathescape, escapeinside=||}
    \newminted[lean4]{tools/lexers/Lean4Lexer.py:Lean4Lexer}{customlexer, mathescape}
  }
  {
    \newminted[mlir]{tools/lexers/MLIRLexer.py:MLIRLexerOnlyOps -x}{mathescape, breaklines, escapeinside=||}
    \newminted[xdsl]{tools/lexers/MLIRLexer.py:MLIRLexer -x}{mathescape, style=murphy}
    \newminted[lean4]{tools/lexers/Lean4Lexer.py:Lean4Lexer -x}{mathescape}
  }
}
\makeatother
\usepackage{xcolor}
\usepackage{soul}

\definecolor{darkGray}{HTML}{353535}
\definecolor{pairedNegOneLightGray}{HTML}{cacaca}
\definecolor{pairedNegTwoDarkGray}{HTML}{827b7b}
\definecolor{pairedOneLightBlue}{HTML}{a6cee3}
\definecolor{pairedTwoDarkBlue}{HTML}{1f78b4}
\definecolor{pairedThreeLightGreen}{HTML}{b2df8a}
\definecolor{pairedFourDarkGreen}{HTML}{33a02c}
\definecolor{pairedFiveLightRed}{HTML}{fb9a99}
\definecolor{pairedSixDarkRed}{HTML}{e31a1c}

\newcommand{\mlirdialect}[1]{%
  \texttt{\textcolor{darkGray}{#1}}%
}
\newcommand{\mlirop}[2]{%
  \texttt{\textcolor{darkGray}{#1.}{\bfseries #2}}%
}

\definecolor{mygray}{gray}{0.6}

\newcommand{\gray}[1]{\textcolor{mygray}{#1}}

\DeclareAcronym{ir}{
  short=IR,
  long=intermediate representation,
}
\DeclareAcronym{ssa}{
  short=SSA,
  long=static single assignment,
}
\DeclareAcronym{cse}{
  short=CSE,
  long=common subexpression elimination,
}
\DeclareAcronym{aegraph}{
 short=ægraph,
 long=acyclic e-graph,
}
\DeclareAcronym{cfg}{
    short=CFG,
    long=control flow graph,
}
\begin{document}

\title{\mlirdialect{eqsat}: An Equality Saturation Dialect for Non-destructive Rewriting}

\author{Jules Merckx}
\email{jules.merckx@ugent.be}
\affiliation{%
  \institution{Ghent University}
  \city{Ghent}
  \country{Belgium}
}

\author{Alexandre Lopoukhine}
\affiliation{%
 \institution{University of Cambridge}
 \city{Cambridge}
 \country{UK}}

\author{Samuel Coward}
\affiliation{%
 \institution{University of Cambridge}
 \city{Cambridge}
 \country{UK}}

\author{Jianyi Cheng}
\affiliation{%
 \institution{University of Edinburgh}
 \city{Edinburgh}
 \country{UK}}

\author{Bjorn De Sutter}
\affiliation{%
  \institution{Ghent University}
  \city{Ghent}
  \country{Belgium}
}

\author{Tobias Grosser}
\affiliation{%
 \institution{University of Cambridge}
 \city{Cambridge}
 \country{UK}}

\renewcommand{\shortauthors}{Merckx et al.}

\begin{abstract}
With recent algorithmic improvements and easy-to-use libraries, equality saturation is being picked up for hardware design, program synthesis, theorem proving, program optimization, and more.
Existing work on using equality saturation for program optimization makes use of external equality saturation libraries such as egg, typically generating a single optimized expression. In the context of a compiler, such an approach uses equality saturation to replace a small number of passes. In this work, we propose an alternative approach that represents equality saturation natively in the compiler's intermediate representation, facilitating the application of constructive compiler passes that maintain the e-graph state throughout the compilation flow.
We take LLVM's MLIR framework and propose a new MLIR dialect named \mlirdialect{eqsat} that represents e-graphs in MLIR code.
This not only provides opportunities to rethink e-matching and extraction techniques by orchestrating existing MLIR passes, such as common subexpression elimination, but also avoids translation overhead between the chosen e-graph library and MLIR.
Our \mlirdialect{eqsat} \ac{ir} allows programmers to apply equality saturation on arbitrary domain-specific \acs{ir}s using the same flow as other compiler transformations in MLIR.
\end{abstract}

\maketitle

\section{Introduction}
To date equality saturation has largely been used outside the compilation flow or has replaced a single compiler pass, with just one work exploring a deep integration~\cite{cranelift}.
Most works that leverage equality saturation for program optimization develop custom tools~\cite{Panchekha2015AutomaticallyExpressions, Coward2024ROVER:Rewriting} built on top of existing equality saturation libraries~\cite{Willsey2021Egg:Saturation,Zhang2023BetterSaturation}. More recent work has now started to explore the integration of equality saturation in general-purpose compiler frameworks~\cite{dialegg2025,Cheng2024SEER:MLIR}. These approaches develop an extensible translation layer between the compiler ecosystem~\cite{mlir2021} and existing equality saturation library implementations~\cite{Willsey2021Egg:Saturation,Zhang2023BetterSaturation}.
While such an approach offers improved support for non-destructive rewriting of \acf{ir}, it does not fully bridge the gap between equality saturation and compilers.
For one, supporting new \ac{ir} primitives requires additional labor in extending the translation layer between the tools~\cite{dialegg2025}.
More importantly, jumping between compiler and external equality saturation library hampers the ability to keep track of equality information as other compiler passes are applied.

\begin{listing}
    \begin{minted}[escapeinside=||]{py}
|def \textcolor{pairedTwoDarkBlue}{normalize\_probs}(logits):|
    |return \textcolor{pairedSixDarkRed}{softmax}(logits, dim=-1)|
|def \textcolor{pairedFourDarkGreen}{compute\_log\_probs}(probs):|
    |return \textcolor{pairedSixDarkRed}{log}(probs)|
    
|def model\_forward(logits):|
    |probs = \textcolor{pairedTwoDarkBlue}{normalize\_probs}(logits)|
    |log\_probs = \textcolor{pairedFourDarkGreen}{compute\_log\_probs}(probs)|
    |return log\_probs|
    \end{minted}
    \caption{Python code where subsequent calls to {\tt \textcolor{pairedSixDarkRed}{softmax}} and {\tt \textcolor{pairedSixDarkRed}{log}} in {\tt model\_forward} are hidden behind call barriers, preventing a rewrite from taking place.}
    \label{list:inlining}
\end{listing}
Take for example the combination of equality saturation with function call inlining, where the compiler replaces a function call by the code of the function body itself.
High-level functions used by programmers abstract a lot of functionality, and can give rise to interesting rewrite opportunities. Some function calls can for example be rewritten to calls to faster, or more precise implementations.
An example often encountered in code using deep learning libraries, is that of {\tt logsoftmax}, where a call to a softmax function followed by a call to logarithm can be replaced by a call to the more numerically stable {\tt logsoftmax} operation.
\begin{align*}
    {\tt call(\textcolor{pairedSixDarkRed}{log}, call(\textcolor{pairedSixDarkRed}{softmax}, x))} \rightarrow {\tt call(\textcolor{pairedSixDarkRed}{logsoftmax}, x)}
\end{align*}

Existing equality saturation techniques operate on a single function body at a time, potentially with some calls already inlined.
In general, however, there is no guarantee that the correct inlining has been applied to expose rewriting opportunities.
As an example (Listing~\ref{list:inlining}), a function {\tt model\_forward} might subsequently call {\tt \textcolor{pairedTwoDarkBlue}{normalize\_probs}} and {\tt \textcolor{pairedFourDarkGreen}{compute\_log\_probs}} that in turn call {\tt \textcolor{pairedSixDarkRed}{softmax}} and {\tt \textcolor{pairedSixDarkRed}{log}}, respectively. Since those function calls are not inlined, the rewrite opportunity cannot be exploited.
Moreover, inlining a function can lead to \emph{less} rewrite opportunities. When {\tt \textcolor{pairedTwoDarkBlue}{normalize\_probs}} and {\tt \textcolor{pairedFourDarkGreen}{compute\_log\_probs}} are inlined, the rewrite can occur, but if one of both {\tt \textcolor{pairedSixDarkRed}{softmax}} or {\tt \textcolor{pairedSixDarkRed}{log}} is inlined \emph{as well}, the rewrite opportunity vanishes.
In essence, function inlining is subject to a phase-ordering problem similar to the one solved by equality saturation for rewriting.
By bringing equality information from term rewriting to other compiler passes such as function call inlining, new rewriting opportunities are revealed.


In this work, we bring first-class equality saturation to \ac{ir}s based on \ac{ssa}. Our native implementation of equality saturation in an existing compiler ecosystem  maximizes reuse and facilitates switching between equality saturation and destructive rewriting with just one additional compiler pass. While this flexibility may come at the expense of some performance, specifically for e-matching and congruence closure, in practice, e-graph rewriting often does not dominate the overall runtime. For example, common sub-expression aware extraction methods that utilize integer linear program solving can often dominate~\cite{Coward2022FormalProver}.
To represent e-graphs in MLIR, we introduce a new MLIR dialect named \mlirdialect{eqsat}, that directly interfaces with existing MLIR dialects. This enables the off-the-shelf reuse of existing compiler passes to implement some of equality saturation's core algorithms, such as congruence closure.

Our contributions are:
\begin{itemize}
    \item expressing equality saturation directly in a compiler's \ac{ir} via a new compiler \ac{ir},
    \item mapping of equality saturation concepts to existing compiler concepts, 
    \item a framework that maintains the e-graph state across compiler transformations, and
    \item a prototypical open-source implementation of the proposed approach in xDSL, a Python-Native single static assignment-based compiler closely mirroring MLIR. 
\end{itemize}

\section{Background}
Our work takes inspiration from recent developments in equality saturation and builds on compiler infrastructure used in production systems.
We leverage existing rewrite patterns with mutating semantics, and instead apply them non-destructively.

\subsection{Equality Saturation}
Equality saturation is a rewriting technique that applies non-destructive rewrites by keeping track of the original expression alongside transformed ones~\cite{Tate2009EqualityOptimization,Willsey2021Egg:Saturation}.
At the heart of most equality saturation libraries is the e-graph datastructure that consists of e-classes, each a collection of e-nodes~\cite{Nelson1980TechniquesVerification}. Each e-node in a particular e-class represents a function (or literal) that is equivalent to the others.
During equality saturation, rewrite patterns are matched against the e-graph. When a match is found, instead of destructively rewriting a term, the new, equivalent term is added as an e-node in the e-class of the original term.
By applying rewrites non-destructively, there is no risk of running into the phase-ordering problem, where one rewrite renders more interesting rewrites impossible. 
In order to efficiently reason over equalities, the e-graph maintains an equality closure under congruence, meaning that different applications of a function on equivalent terms, yield equivalent terms.

Three prior works have combined equality saturation with a mature compiler framework. The SEER project~\cite{Cheng2024SEER:MLIR}, specifically targeted high-level synthesis, optimizing System-C programs using a combination of high-level software rewrites and low-level circuit rewrites. The second, more general work~\cite{dialegg2025}, developed a framework for representing any internal MLIR dialect in an e-graph, allowing users to define their own equality saturation rewrites. Both works leveraged existing equality saturation libraries, adding translation layers between the two domains.

In contrast, Cranelift~\cite{cranelift}, a mature optimizing compiler and code generator, developed an equality saturation optimizer that reuses much of the infrastructure of their \ac{ir}. This is similar to the approach we present in this paper, so we provide a thorough comparison in Section~\ref{sec:related}.

\subsection{Static Single Assignment \ac{ir} with Regions}

\Ac{ssa} is a property of \ac{ir}s that guarantees that a value is defined exactly once.
Modern compilers~\cite{Lattner2004LLVM,gcc-compiler} leverage this property to simplify analysis and transformations during compilation.
Values are defined as the results of \textit{operations}, or given as arguments to \textit{blocks}.
Operations represent run-time information, such as integer addition, taking a variable number of values as \textit{operands}, and returning a variable number of values as \textit{results}.
Blocks group together a number of operations, either encoding a sequence of operations to be run one after the other, or a cyclic graph of relationships between values with no explicit control flow.
In contrast to textbook \ac{ssa} implementations, which represent values passed into a block from divergent control flow using phi nodes, more recent compilers~\cite{swift,mlir2021,cranelift} instead use block \textit{arguments}.
The flexibility of this data structure has led to its widespread use in modern compilers spanning applications from tensor programs~\cite{bik2022compiler,vasilache2022composable} to digital circuits~\cite{calyxASPLOS2021}.

We use xDSL~\cite{xdsl} to implement equality saturation using \ac{ssa} constructs as defined in MLIR~\cite{mlir2021}.
In MLIR, blocks are nested within \textit{regions}, which are in turn nested within \textit{operations}, allowing users to represent recursive structures of arbitrary depth.
Operation definitions are grouped into user-provided \textit{dialects}, which serve as a name space for related definitions.
In order to provide reusable infrastructure for user-provided constructs, MLIR leverages \textit{interfaces} and \textit{traits} to encode properties and behaviors of operations, to be leveraged by transformations such as dead code elimination or \ac{cse}.

\paragraph{The Pattern Description Language dialect}
The MLIR project contains a number of meta-dialects, such as \mlirdialect{pdl}, which encodes definitions of rewrites on MLIR operations, and \mlirdialect{pdl\_interp}, comprising the actions taken by a state machine when executing the rewrite.
Rewrites defined in \mlirdialect{pdl} are composed of two parts, the first a declarative matching pattern, and the second an imperative rewrite procedure.
A number of these patterns can be lowered together into a single unified state machine, to be executed by an accompanying interpreter defined in MLIR.
The patterns are defined to be applied destructively, meaning that a matched pattern will replace some of the existing \ac{ir}.

\section{E-Graphs as a Compiler \ac{ir}}\label{sec:eqsat_dialect}
To make equality saturation dynamically available throughout the compilation process, we introduce \ac{ir} primitives for modeling the data structures at the core of equality saturation.
Our primitives are a set of \ac{ssa} operations implemented in MLIR. Together, they form our new \mlirdialect{eqsat} dialect.

As an example, we consider the \ac{ir} of the function ${\tt a * 2}$ (Listing~\ref{list:eqsat_conversion} - top), which takes one argument and multiplies it by two.
We can express this function as a simple expression tree consisting of a multiplication node with two children, \texttt{a} and \texttt{2}.
We can then turn this \ac{ir} into a trivial e-graph (where each class has just one element) by introducing equality class operations for each result (Listing~\ref{list:eqsat_conversion} - bottom).
In particular, we introduce three new \mlirop{eqsat}{eclass} operations that use the argument \texttt{\%a} as well as the results \texttt{\%two} and \texttt{\%res} and return corresponding e-classes, each containing exactly one element. We subsequently update the
compute operation \mlirop{arith}{muli} to use the newly created equality class result values as its inputs.
As the e-graph is being rewritten, circular references may be introduced, which may be disallowed by the parent operation.
To preserve correctness, we embed our e-graph in an \mlirop{eqsat}{egraph} operation, which encapsulates the e-graph, preserving the validity of the rest of the \ac{ir}.
The resulting \ac{ir} models an e-graph (at the right) with three equality classes, each visualized as a dotted frame.

We can now introduce new equalities into our \ac{ir}. For example, the multiplication in our program is equal to a more efficient left shift.
After applying the corresponding rewrite ${\tt x*2}\rightarrow {\tt x << 1}$ to our \ac{ir} (Listing~\ref{list:eqsat_equality}), a new bitshift operation as well as the necessary constant operation have been added to the code, and the result of the bitshift has been added as an operand to the same \mlirop{eqsat}{eclass} operation that already referenced the multiplication result.  This new \ac{ir} now corresponds to
an e-graph where the results of multiplication and left shift are
members of the same e-class (Listing~\ref{list:eqsat_equality} - right). While visualized differently, the reader may observe that each \mlirop{eqsat}{eclass} operation corresponds to exactly one e-class in the egraph, while the edges of the e-graph correspond to use-def edges that connect the results of \mlirop{eqsat}{eclass} operations with the arithmetic operations \mlirop{arith}{muli} and \mlirop{arith}{shli}. Hence, an e-graph can be trivially embedded into an \ac{ssa}-based compiler \ac{ir}.
\begin{listing}
\begin{minipage}[t]{0.6\linewidth}
\small
\begin{xdsl}
|\gray{func.func @f(}\textcolor{pairedOneLightBlue}{\%a}\gray{ : i64) -> i64 \{}|
  |\textcolor{pairedThreeLightGreen}{\%two = arith.constant 2 }\gray{: i64}|
  |\textcolor{pairedFiveLightRed}{\%res = arith.muli }\textcolor{pairedOneLightBlue}{\%a}\gray{, }\textcolor{pairedThreeLightGreen}{\%two}\gray{ : i64}|
  |\gray{func.return }\textcolor{pairedFiveLightRed}{\%res}\gray{ : i64}|
|\gray{\}}|
\end{xdsl}
\end{minipage}\hfill
\begin{minipage}[t]{0.2\linewidth}
  \vspace{-1mm}
    \centering
    \includegraphics{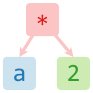}
\end{minipage}

\vspace{5mm}
\begin{minipage}[t]{0.6\linewidth}
\small
\begin{xdsl}
|\gray{func.func @f(}\textcolor{pairedOneLightBlue}{\%a}\gray{ : i64) -> i64 \{}|
  |\textcolor{pairedThreeLightGreen}{\%two = arith.constant 2 }\gray{: i64}|
  |\gray{\%graph\_res = eqsat.egraph -> i64 \{}|
    |\textcolor{pairedFourDarkGreen}{\%c\_two = eqsat.eclass }\textcolor{pairedThreeLightGreen}{\%two}\gray{ : i64}|
    |\textcolor{pairedTwoDarkBlue}{\%c\_a = eqsat.eclass }\textcolor{pairedOneLightBlue}{\%a}\gray{ : i64}|
    |\textcolor{pairedFiveLightRed}{\%res = arith.muli }\textcolor{pairedTwoDarkBlue}{\%c\_a}\gray{, }\textcolor{pairedFourDarkGreen}{\%c\_two}\gray{ : i64}|
    |\textcolor{pairedSixDarkRed}{\%c\_res = eqsat.eclass }\textcolor{pairedFiveLightRed}{\%res}\gray{ : i64}|
    |\gray{eqsat.yield }\textcolor{pairedSixDarkRed}{\%c\_res}\gray{ : i64}|
  |\gray{\}}|
  |\gray{func.return 
|\gray{\}}|
\end{xdsl}
\end{minipage}\hfill
\begin{minipage}[t]{0.2\linewidth}
  \vspace{5mm}
    \centering
    \includegraphics{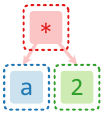}
\end{minipage}
\caption{We implemented a compiler pass that inserts operations from our \mlirdialect{eqsat} dialect in order to represent e-graphs within MLIR code directly.}
\label{list:eqsat_conversion}
\end{listing}

Concretely, our \mlirdialect{eqsat} dialect consists of three operations which in combination with the use-def information offered by an \ac{ssa}-based compiler \ac{ir} are sufficient to represent e-graphs and carry out equality saturation:
\begin{itemize}
    \item The \mlirop{eqsat}{eclass} operation takes one or more values (analogous to e-nodes), and produces a single result.
    \item The \mlirop{eqsat}{egraph} operation encompasses a piece of code on which equality saturation can be executed. All \mlirop{eqsat}{eclass} operations must be contained within an \mlirop{eqsat}{egraph} operation. Operations within this operation's region can access values defined outside of it, but not the other way around.
    \item The \mlirop{eqsat}{yield} operation is a terminator that closes off an e-graph. It takes as operands the \mlirop{eqsat}{eclass} results that are exposed by the \mlirop{eqsat}{egraph} operation to the rest of the program.
\end{itemize}
\begin{listing}
\begin{minipage}[t]{0.7\linewidth}
\small
\begin{xdsl}
|\textcolor{black}{\%one = arith.constant 1}|
|\textcolor{black}{\%c\_one = eqsat.eclass \%one}|
|\textcolor{pairedFourDarkGreen}{\%c\_two = eqsat.eclass }\textcolor{pairedThreeLightGreen}{\%two}|
|\textcolor{pairedTwoDarkBlue}{\%c\_a = eqsat.eclass }\textcolor{pairedOneLightBlue}{\%a}|
|\textcolor{pairedFiveLightRed}{\%res = arith.muli }\textcolor{pairedTwoDarkBlue}{\%c\_a}\gray{, }\textcolor{pairedFourDarkGreen}{\%c\_two}|
|\textcolor{black}{\%res1 = arith.shli }\textcolor{pairedTwoDarkBlue}{\%c\_a}\gray{, }\textcolor{black}{\%c\_one}|
|\textcolor{pairedSixDarkRed}{\%c\_res = eqsat.eclass }\textcolor{pairedFiveLightRed}{\%res}\gray{, }\textcolor{black}{\%res1}|
|\gray{eqsat.yield }\textcolor{pairedSixDarkRed}{\%c\_res}|
\end{xdsl}
\end{minipage}\hfill
\begin{minipage}[t]{0.3\linewidth}
  \vspace{-1mm}
    \centering
    \includegraphics{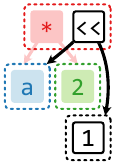}
\end{minipage}
\caption{Contents of the \mlirop{eqsat}{egraph} operation from Listing~\ref{list:eqsat_conversion}, and corresponding e-graph visualization, after the rewrite ${\tt x*2}\rightarrow {\tt x << 1}$ has been applied.}
\label{list:eqsat_equality}
\end{listing}

\paragraph{Cycles}
Depending on the equality rules used during equality saturation, cycles can appear in the e-graph.
For example, applying the rule ${\tt a + 0} \rightarrow {\tt a}$ on an e-graph containing the expression ${\tt a + 0}$ introduces a cycle as illustrated in Listing~\ref{fig:cycle}.
In typical \ac{ssa}-based \ac{ir}s, value uses can only occur after their definition.
MLIR, however, supports the concept of graph regions, where this restriction is lifted.
The region encompassed by an \mlirop{eqsat}{egraph} operation is such a graph region, allowing cycles in the use-def chain to occur.
\begin{listing}
    \centering
\begin{minipage}[t]{0.2\linewidth}
    \vspace{0pt}
    \centering
    \includegraphics{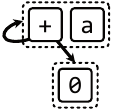}
\end{minipage}\hfill
\begin{minipage}[t]{0.7\linewidth}
    \vspace{4mm}
    \begin{xdsl}
    \end{xdsl}
\end{minipage}
    \caption{Cycles in e-graphs can be represented compactly in MLIR \ac{ir} by using graph-regions.}
    \label{fig:cycle}
\end{listing}

\paragraph{Control Flow}
By virtue of MLIR's region-based IR, there are dialects that can be used to represent control flow such as if-else-statements or for loops in a structured manner.
For example through the use of the \mlirdialect{scf} dialect, which offers \mlirop{scf}{for}, \mlirop{scf}{if}, and other operations.
In contrast to simple arithmetic operations, these control flow operations carry a region containing the control flow body, allowing the \acp{ir} to express programs with control flow without the use of basic blocks and phi-nodes.
The presence of these nested control flow regions does not hinder equality saturation but rather allows rewrites to naturally occur across control flow.
For example, values defined outside of a for loop are still accessible within the loop while the inverse is not true (Listing~\ref{list:control_flow}).
Simply inspecting the graphical e-graph structure does not reveal these scope constraints, they are encoded only in the \ac{ir}.

\begin{listing}
\begin{minipage}{0.6\linewidth}
\small
\begin{xdsl}
|\textcolor{pairedTwoDarkBlue}{\%s = scf.for }\textcolor{pairedFourDarkGreen}{\%i}\textcolor{pairedTwoDarkBlue}{ = \%lb to \%ub}|
|     \textcolor{pairedTwoDarkBlue}{iter\_args(}\textcolor{pairedFourDarkGreen}{\%s}\textcolor{pairedTwoDarkBlue}{ = \%s\_0) \{}|
|  \textcolor{pairedFourDarkGreen}{\%x = memref.load }\textcolor{pairedTwoDarkBlue}{\%a}\textcolor{pairedFourDarkGreen}{[\%i]}|
|  \textcolor{pairedFourDarkGreen}{\%term = arith.mulf \%x, }\textcolor{pairedTwoDarkBlue}{\%two}|
|  \textcolor{pairedFourDarkGreen}{\%s\_new = arith.addf \%s, \%term}|
|  \textcolor{pairedFourDarkGreen}{scf.yield \%s\_new}|
|\textcolor{pairedTwoDarkBlue}{\}}|
\end{xdsl}
\end{minipage}\hfill
\begin{minipage}{0.38\linewidth}
  \vspace{0pt}
    \centering
    \includegraphics{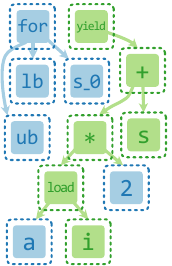}
\end{minipage}
\caption{Example of an \mlirop{scf}{for} operation, and the corresponding e-graph. \mlirop{eqsat}{eclass} operations have been left out in this example for clarity. Operations \emph{\textcolor{pairedFourDarkGreen}{inside}} the loop body can access values from \emph{\textcolor{pairedTwoDarkBlue}{outside}}.
}
\label{list:control_flow}

\end{listing}


\pagebreak
\section{Rebuilding}
\begin{listing}
\begin{minipage}[t]{0.5\linewidth}
    \vspace{2mm}
    \small
    \begin{xdsl}
    \end{xdsl}
    \vspace{8mm}
    \begin{xdsl}
    \end{xdsl}
    \vspace{11mm}
    \begin{xdsl}
    \end{xdsl}
\end{minipage}\hfill
\begin{minipage}[t]{0.45\linewidth}
  \vspace{0pt}
    \centering
    \includegraphics{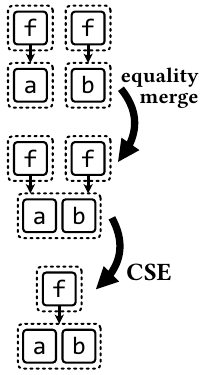}
\end{minipage}
\caption{By moving equality saturation primitives into the compiler, CSE subsumes egraph rebuilding.}
\label{fig:rebuilding}
\end{listing}
At the core of equality saturation is the congruence invariant that ensures that all equalities in an e-graph are closed under congruence.
To conserve this invariant more efficiently, Willsey et al.~\cite{Willsey2021Egg:Saturation} proposed an explicit e-graph rebuilding step, where the merging of parent e-classes is carried out for all the new equalities that have been inserted after matching all patterns, instead of immediately after each single insertion.

We make the observation that, by embedding e-classes as operations in an \ac{ssa}-based \ac{ir}, applying \ac{cse} maintains this invariant as well.
As an example, take a function {\tt f} being applied twice with different e-class operands containing {\tt a} and {\tt b}, respectively (Listing~\ref{fig:rebuilding}).
When an equality $a \leftrightarrow b$ is introduced, the equality saturation pass merges the equivalent e-classes. All uses of the original e-classes now use this merged e-class instead.
At this point, the congruence invariant is not satisfied, as the two e-classes containing {\tt f} are equivalent because they refer to the same e-class operand.
Applying \ac{cse} on the MLIR code removes the duplicated application of {\tt f}, thus restoring the invariant.

By utilizing a standard \ac{cse} pass as provided by MLIR, the implementation of equality saturation is made considerably simpler.
Importantly, the \ac{cse} pass itself was implemented without equality saturation in mind.
Compared to the full e-graph rebuilding algorithm, however, \ac{cse} can be less efficient as it is applied on the whole e-graph instead of incrementally, on newly introduced equalities.
In the future, it might be possible to leverage incremental \ac{cse}\footnote{\url{https://mlir.llvm.org/doxygen/CSE_8cpp_source.html}} for more efficient e-graph rebuilding.


\section{Grafting E-Matching}
In order to find and apply rewrites on an e-graph, e-matching needs to be performed.
Here, the e-graph is searched, verifying if any of the provided patterns are matched in the equivalence structure.
With the \mlirdialect{pdl} dialect it is possible to describe complex patterns.
Static types of operation results and operands can be matched in order to describe complex type constraints (Listing~\ref{list:pdl}).
Additionally, \emph{multi-patterns}, where multiple expressions are matched and rewritten together, are trivially expressed in \mlirdialect{pdl}'s declarative format.
\begin{listing}
\small
\begin{xdsl}
pdl.pattern : benefit(1) {
         -> (
       ) -> (
  pdl.rewrite 
    pdl.replace 
  }
}
\end{xdsl}
\caption{A declarative rewrite pattern written in MLIR's \mlirdialect{pdl} dialect to rewrite $a + 0$ into $a$.}
\label{list:pdl}
\end{listing}

Typically, \mlirdialect{pdl} rewrite patterns are first lowered into operations from the \mlirdialect{pdl\_interp} dialect.
This dialect consists of lower level, imperative matching operations, making it simpler for an interpreter to match and rewrite. Listing~\ref{list:pdl_interp} shows the \mlirdialect{pdl\_interp} code that is obtained by lowering the \mlirdialect{pdl} code from Listing~\ref{list:pdl}.

The existing lowering pass from \mlirdialect{pdl} to \mlirdialect{pdl\_interp} also combines multiple patterns into one search routine, reusing information used for different patterns and bailing out early as soon as none of the patterns can be matched anymore.

\begin{listing}
\footnotesize
\begin{xdsl}
pdl_interp.func @matcher(
  pdl_interp.is_not_null 
^bb1:
  pdl_interp.finalize
^bb2:
  pdl_interp.check_operation_name of 
    -> ^bb3, ^bb1
^bb3:
  pdl_interp.check_operand_count of 
\end{xdsl}
\vspace{-5pt}
\begin{center}
$\vdots$
\end{center}

\caption{Part of the result after lowering the \mlirdialect{pdl} rewrite pattern from Listing~\ref{list:pdl} to the \mlirdialect{pdl\_interp} dialect. This imperative code contains simple instructions and primitive control flow to match one or more patterns.}
\label{list:pdl_interp}
\end{listing}

In order to reuse this existing rewrite infrastructure for matching patterns in the \ac{ir} that has been extended with \mlirdialect{eqsat} operations, we have implemented an alternative interpreter over \mlirdialect{pdl\_interp} operations that takes into account the extra indirections caused by \mlirop{eqsat}{eclass} operations. Most importantly, this interpreter is responsible for backtracking and trying out all possible values in an e-class, as is done by most equality saturation frameworks, and described by De Moura et al.~\cite{DeMoura2007EfficientSolvers}. 

\begin{figure}
    \centering
    \vspace{-3mm}
    \includegraphics{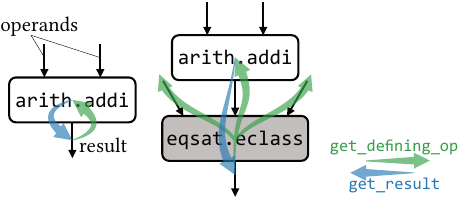}
    \vspace{-3mm}
    \caption{(left) In MLIR's default, destructive rewriting, \mlirop{pdl\_interp}{get\_result} and \mlirop{pdl\_interp}{get\_defining\_op} are each other's inverse. (right) In equality saturation, this is not the case because each value comes from one of multiple equivalent operations.}
    \label{fig:pdl_interp}
\end{figure}

As it turns out, for equality saturation, most \mlirdialect{pdl\_interp} operations can be interpreted exactly the same as for classical rewriting, with the exception of \mlirop{pdl\_interp}{get\_result} and \mlirop{pdl\_interp}{get\_defining\_op}, as these operations go back and forth between an operation and its result.  
Similarly, the behavior of \mlirop{pdl\_interp}{create\_operation}, and \mlirop{pdl\_interp}{replace} needs to be adapted to take into account the e-class operations.
The \mlirop{pdl\_interp}{get\_result} operation takes an operation and returns its result value (Figure~\ref{fig:pdl_interp}). Using our \mlirdialect{eqsat} dialect, the result of all operations in the e-graph are \mlirop{eqsat}{eclass} operations, which is not what the existing pattern matching code expects.
Instead, our interpreter digs one level deeper, returning the result of the \mlirop{eqsat}{eclass} operation.

Similarly, for \mlirop{pdl\_interp}{get\_defining\_op}, an operation that returns the operation that defines a particular value, the link from result to defining operation is interrupted by an \mlirop{eqsat}{eclass} operation (Figure~\ref{fig:pdl_interp}).
To handle this, the interpreter keeps track of each \mlirop{pdl\_interp}{get\_defining\_op} operation. And when a pattern match fails, the interpreter goes back to the latest instance and retries matching with the next operand of the \mlirop{eqsat}{eclass} operation.

For \mlirop{pdl\_interp}{create\_operation}, instead of blindly creating and inserting a new operation, the interpreter will now first verify if an identical operation already exists in the program and use that one.
This serves the same purpose as the use of hashconsing in typical e-graph libraries.
Lastly, \mlirop{pdl\_interp}{replace} now does not remove the operation being replaced, but rather inserts the replacement values in the correct e-class.

By again depending on existing compiler infrastructure to implement part of equality saturation, development burden is greatly reduced.
Additionally, since MLIR combines all patterns in a single matching routine, large rulesets with overlapping subpatterns can more efficiently be pruned compared to most existing equality saturation frameworks. For example, in egg patterns are matched one by one, in a top-down manner. Another approach looks at pattern matching on e-graphs from a database perspective~\cite{zhang2022relational}, converting e-graphs into a relational database and viewing pattern matching as a relational join.
Similar to this approach, the generated MLIR pattern matcher is able to search for patterns top-down, bottom-up, or a combination thereof using existing pattern matching infrastructure. 


\section{Future Work}
Our first step will be to explore mechanisms for combining multiple cost models, useful in cases where performance must be traded off with floating-point accuracy~\cite{Panchekha2015AutomaticallyExpressions,chassis2025asplos} in linear algebra micro-kernel compilation~\cite{riscV2025CGO}.
Combining cost models is also inevitable when the program being rewritten is expressed in terms of operations in multiple dialects, with associated cost being computed by separate cost models. Furthermore, having equivalent program representations at different levels of abstraction provides the opportunity to combine analyses across abstraction levels~\cite{Coward2023CombiningInterpretation}. 
In circuit design, for example (Listing~\ref{list:sign_extension}), the \mlirdialect{moore} dialect uses just one operation, while the \mlirdialect{comb} dialect utilizes three operations to achieve the same result.
Naturally, the single operator is simpler to analyze, say via an interval analysis, while the three-operator implementation is easier to lower into real hardware.

\begin{listing}
\small
\begin{xdsl}

\end{xdsl}
\caption{Equivalent representations of sign-extension from a two-bit value to a four-bit value.}
\label{list:sign_extension}
\end{listing}
As discussed (Section~\ref{sec:eqsat_dialect}), region-based control flow operations do not inhibit equality saturation. Currently, however, the \mlirdialect{pdl} dialect cannot be used to match regions of operations. This means that, while it is possible to match code in regions, it is not yet possible to match complete control flow operations and rewrite those. In the future, allowing this could open up doors to not only rewrite code in the presence of control flow, but also rewrite control flow operations themselves.

\pagebreak
\section{Related Work}\label{sec:related}
The idea of embedding equality saturation in a compiler has been explored before with Cranelift's \acp{aegraph}~\cite{cranelift}.
Although similar to our work, we have made a number of different design decisions that lead to distinct capabilities.

Firstly, functions in Cranelift's \ac{ir} contain \acp{cfg}, possibly consisting of multiple basic blocks and unstructured control flow.
The presence of a \ac{cfg} complicates building an e-graph representation, as phi-nodes (or block arguments) complicate def-use dependencies by making them conditional on control flow.
To resolve this, Cranelift introduces the concept of a CFG skeleton, a data structure storing the fixed \ac{cfg} such that the function can be reconstructed from the e-graph representation.
The CFG skeleton has the downside of prohibiting control flow rewrites, meaning that rewrites that span multiple basic blocks can occur, but that the structure of the basic blocks and control flow is fixed.

By building on MLIR, we instead target code using structured control flow constructs. Here, every function consists of a single basic block, and control flow is captured by operations containing nested code regions.
Conceptually, this can allow control flow to be rewritten just the same as other operations.

Secondly, Cranelift's IR is strictly SSA, and does not have the concept of graph regions making it less straightforward to represent cyclic e-graphs in IR directly.
Instead, Cranelift rewriting runs in a single pass, applying multiple rewrites eagerly the moment each instruction is added to the e-graph.
By leveraging graph regions, we are able to represent cyclic e-graphs, allowing us to execute full equality saturation in our framework.

Lastly, by targeting MLIR, instead of a lower level \ac{ir} such as the one found in Cranelift, we open up the possibility of rewriting in many different domains, and across different abstraction levels.

\section{Conclusion}
Existing work using equality saturation for code optimization mostly does so by using external libraries. This prevents combining the advantages of equality saturation with existing compiler analyses and transformations, and requires additional work in order to support new program constructs.
By bringing equality saturation to the compiler in the form of \ac{ir} primitives, the fundamental barrier between equality saturation and compiler passes is lifted.
We have shown that existing compiler passes such as \ac{cse}, and compiler infrastructure such as MLIR's \mlirdialect{pdl} dialect can be harnessed to more easily implement equality saturation.
Furthermore, this approach opens up many new opportunities for using equality saturation to further enhance existing compiler passes.

\begin{acks}
This work has received funding from the European Union’s Horizon EUROPE research and innovation program under grant agreement no. 101070375 (CONVOLVE).
\end{acks}

\bibliography{references}


\end{document}